\documentstyle[aps,preprint]{revtex}
\begin{document}
\title{
Off-Diagonal Long Range Order and Scaling
in a Disordered Quantum Hall System
}
\author{S. L. Sondhi}
\address{
Department of Physics, University of Illinois at Urbana-Champaign, Urbana,
Illinois 61801-3080
}
\author{M. P. Gelfand}
\address{
Department of Physics, Colorado State University, Fort Collins, Colorado 80523
}
\maketitle
\begin{abstract}
We have numerically studied the bosonic off-diagonal long range order,
introduced by Read to describe the ordering in ideal quantum Hall states,
for noninteracting electrons in random potentials
confined to the lowest Landau level.
We find that it also describes the ordering in disordered quantum Hall states:
the proposed order parameter vanishes in the disordered ($\sigma_{xy}=0$)
phase and increases continuously from zero in the ordered ($\sigma_{xy}=e^2/h$)
phase.
We study the scaling of the order parameter and find that it is consistent
with that of the one-electron Green's function.

\end{abstract}
\pacs{73.40.Hm, 71.30.+h, 71.27.+a}
\narrowtext
The quantum Hall effect (QHE) is a consequence of novel correlated states that
arise in a two dimensional electron gas placed in a transverse magnetic field
\cite{ref:bible}. In ideal, {\it i.e.} translationally invariant, systems these
states exist at isolated filling factors at which the system is incompressible.
In systems with impurities
these states broaden into phases --- ranges of filling factor
($\nu$) which exhibit the same transport properties as the parent ideal
states --- thus giving rise to the characteristic plateaux structure of the
QHE.
The nature of the ordering in the ideal states was elucidated by Girvin and
MacDonald (GM) \cite{ref:gm} and later by Zhang, Hansson and Kivelson
(ZHK) \cite{ref:zhk} and by Read \cite{ref:read}, who
showed that the Laughlin states could be viewed as condensates of composite
bosons consisting of electrons that ``carry'' flux. The GM/ZHK formulation is
distinct from that of Read and there is no proof that they are equivalent. In
this paper we will be concerned, for reasons of computational convenience,
solely with Read's formulation; we
comment briefly on the GM case at the end.

In this work we address the following questions: Is Read's
bosonic off-diagonal long range order (ODLRO) a property of real, dirty
quantum Hall systems, i.e. is it non-zero in the entire phase descended from
an ideal state and does it vanish outside its boundaries?
(That the ODLRO vanishes for a clean system when the Laughlin states are
destabilized by varying the electron-electron interaction was shown
already by Rezayi and Haldane \cite{ref:rh-read}.)
We emphasize that at issue here is whether the ODLRO can serve as a sufficient
characterization of the real systems that exhibit the QHE; note that
incompressibility is lost when disorder is introduced into the system.
Anticipating that the ODLRO does survive the introduction of disorder,
we are led to the derivative questions of the critical
behavior of the Read bosons near the transition between neighboring Hall
plateaux, ``seen'' by them as a superfluid-insulator
transition \cite{ref:lkz}, and its relationship to the conventional measures
of delocalization.

We will present likely answers to these questions based on numerical
studies of non-interacting electrons subject to random
potentials which are
confined to the lowest Landau level (LLL) --- a problem for which the
localization properties of single-particle states have been investigated
extensively \cite{ref:locrefs}. This choice may seem perplexing on account
of the historical association of the ODLRO with the {\em fractional}
states, but we remind the reader that the integer Hall state at $\nu=1$, the
filled LLL, is just the first state in the Laughlin sequence $1,1/3,1/5 \ldots$
and does exhibit ODLRO; more generally, there is no distinction between the
integer QHE and fractional QHE in this regard and we expect our qualitative
results to hold quite generally.
We note that even for non-interacting electrons the ODLRO is a property of
the {\em many-body} state, and is not a one-electron quantity averaged over
the occupied single-particle states.

Most of the calculations are done on the
sphere \cite{ref:sphere} using a density per flux quantum, $\rho_{\rm i}$, of
delta-function scatterers, and a modified
version of Read's operator introduced by Rezayi and Haldane \cite{ref:rh-read}.
For a sphere
containing $N_\phi=2S +1$ flux quanta the single-particle states can be
labeled by the eigenvalues, $m$, of $L_z$ which run from $-S$ to $S$.
If $L_z$ generates rotations about the point $\bf x$, then
Read's operator for the $\nu=1$ state \cite{ref:fn1} takes the form
\begin{equation}
\phi^{\dagger}_{R}({\bf x};S)=
c^{\dagger}_{m=S+1/2}\, F({\bf x}; S).
\end{equation}
Here $F({\bf x};S)$ is the flux insertion operator at $\bf x$,
which replaces a state
of the system with $N_\phi$ flux quanta and occupations $n_{N_\phi}(m)$ by
a state of the system with $N_\phi+1$ flux quanta and occupations
$n_{N_\phi+1}(m-1/2)=n_{N_\phi}(m)$.
Note that the electron creation operator $c^{\dagger}$ acts on the states
of the $N_\phi+1$ flux system. The intuitive content of this definition
is more easily understood in the planar disk geometry, where the action of
$\phi^{\dagger}_{R}({ \bf 0})$
is the insertion of one flux quantum at the origin causing each of the
single-particle
states to move outwards into its neighbor and the subsequent injection of an
electron into the central Gaussian orbital. At $\nu=1$ it
is clear that this operation takes the $N$ particle ground state
to the $N+1$ particle ground state, much as the operation of the field operator
on the $N$ particle ground state of a superfluid produces a state with an
$O(1)$ overlap with the $N+1$ particle ground state. For dirty systems,
it is not immediately evident that (in obvious notation),
$\langle N+1| \phi^{\dagger}_{R}({\bf x})|N \rangle$ continues to be $O(1)$
for all $\nu$ at which $\sigma_{xy}=e^2/h$ or that it vanishes when
$\sigma_{xy}=0$.
There is, however, a suggestive connection
of $\phi^{\dagger}_{R}$ with the flux insertion in Laughlin's
gauge argument \cite{ref:laugh-gauge}. The latter can be interpreted
as the statement that flux insertion followed by transferring an electron
between edges has {\em no} effect on the ground state in the $\nu=1$
phase \cite{ref:thouless}, but it is important to note that
in the gauge argument the electrons in the localized states are unaffected
by adiabatic flux insertion while in the action of $\phi^{\dagger}_{R}$
{\em all} the electrons are affected.
We note that, in the gauge argument, the inertness of the localized electrons
is essential in order to get a quantized $\sigma_{xy}$. Consequently, and
in contrast, the expectation value of $\phi^{\dagger}_R$ is not quantized.

The principal numerical results concern the disorder-averaged absolute value of
the
antipodal correlator:
\begin{equation}
G_R(\nu,S)= [|\,\langle\phi_R(\theta=0) \phi^{\dagger}_{R}(\theta=\pi)
\rangle|\,]_{\rm dis},
\label{eq:opdef}
\end{equation}
where $\theta=0,\pi$ denote the north and south poles, respectively, and the
expectation value $\langle\rangle$ is evaluated in the $N_e$-electron
ground state. (Note that the filling factor $\nu=N_e/N_\phi$.)  We find:

(1) In the thermodynamic limit ($S\to\infty$),
for $\nu \le 1/2$, $G_R(\nu,S) \rightarrow 0$,
while for $\nu > 1/2$,
$G_R(\nu,S)\rightarrow m^2_R(\nu) \ne 0$, so that
$\phi^\dagger_R$ defines a legitimate
order parameter for the dirty system. As seen from Fig.~\ref{fig1}, $m^2_R$
vanishes continuously on approaching $\nu=1/2$ from above.
The critical point is at $\nu=1/2$ because
the random potentials employed were strictly particle-hole symmetric.
We have also calculated the disorder-averaged absolute value of the
order parameter matrix element, $M_R = [|\langle N_e+1|\phi_R^\dagger
(\theta=0)|N_e\rangle|]_{\rm dis}$, and
checked that it behaves as expected. The antipodal correlator is, however,
easier to calculate and vanishes more rapidly (exponentially with $S$
rather than as $1/S$) in the low-$\nu$ phase, so it was investigated
more thoroughly.  For other practical reasons, we have focused
on the sphere, rather than the disk or torus geometries.

Previous work on destruction of the ODLRO in the ideal states
\cite{ref:gm,ref:read,ref:rh-read} identified the unbinding of zeroes of the
wavefunction from the particles as the relevant mechanism.
In our problem there is always one zero per particle due to fermi statistics;
the ODLRO is destroyed instead by the delocalization of extra zeroes
(quasiholes) introduced by varying $\nu$. Heuristically, the variation of $m_R$
with $\nu$ appears to reflect a geometrical property of the states, namely
the extent of a ``percolating'' cluster of $\nu=1$ liquid. The quantized
Hall conductance, however, is sensitive only to the existence of the cluster.

(2) For $1/2 < \nu < 1$ one expects a state in which randomly localized
quasiholes
are interspersed with a $\nu=1$ condensate. In the bosonic description,
this state resembles a vortex glass \cite{ref:vortglass} where the phase
of the order parameter at a given point in space fluctuates randomly
between disorder configurations on account of fluctuations in the locations
of the vortices (quasiholes). We have confirmed that {\em not\/} taking the
absolute value in Eq.~(\ref{eq:opdef}) causes the disorder average to vanish
at all filling factors $0<\nu<1$.

(3) The data in the critical region admit a finite-size scaling
analysis consistent with the critical behavior found in previous
studies of one-electron measures of localization.
The appropriate scaling ansatz is
\begin{equation}
G_R \sim S^{-\eta/2} g[(\nu-\textstyle{1\over2})S^x],
\label{eq:fss}
\end{equation}
where $x=1/(2\nu_\xi)$ and $\nu_\xi$ is the correlation length exponent.
Note that the antipodal distance varies as the square root of $S$.
We estimate $\nu_\xi>2$ and $\eta\simeq 1.6$; consequently $m^2_R(\nu) \sim
(\nu -1/2)^{2 \beta}$ where $2 \beta = \nu_\xi \eta > 3.2$.
However, the analysis
is complicated by a slow crossover, apparently reflecting a weakly
irrelevant operator.  The analysis is described below.

(4) In order to further test the basic question of the order parameters for
dirty quantum Hall systems, we calculated the
antipodal correlations for
the $\nu=1/3$ order parameter ({\it i.e.}, insertion of three flux quanta
at a point followed creation of an electron there) in the noninteracting
system.  For finite $S$ (and at $\rho_{\rm i}=4$),
the correlations are nonzero, with a maximum
at a filling somewhat less than 1/3.  The maximum value of
the correlation function decays rapidly as the system size increases,
perhaps as $1/S^3$.  These results support the notion that there
is a unique bosonic order parameter associated with each quantum Hall phase.

We now turn to the details of the scaling analysis for $G_R$.
We have data for $G_R$, as well as the equal-time one-electron Green's
function,
at $\rho_{\rm i}=2$, 4, 8, 16 and for system sizes up to $2S=320$.
At any given $\rho_{\rm i}$, Eq.~(\ref{eq:fss}) describes the data
well but the estimated $\nu_\xi$ (but not $\eta$) drifts substantially
with the impurity concentration.

Let us focus on the $\rho_{\rm i}=4$ data for the moment.
The band-center data exhibit systematic deviations from scaling
at the smaller system sizes, see Fig.~2; we estimate $\eta/2=0.78(2)$ based
on the data for $2S \geq 120$, but that value may be too low \cite{ref:fn3}.
Likewise, a log-log plot of $dG_R/d\nu$ at $\nu=1/2$ (determined by fitting,
as were the band-center values themselves) versus $2S$ exhibits curvature.
However, the ratio of the slope to the value scales nicely,
as shown in Fig.~2, leading to an estimate
of $x=0.344(10)$, corresponding to $\nu_\xi=1.45(4)$.  The data collapse in the
corresponding scaling plot, Fig.~3, is evident.
This estimate of $\nu_\xi$ above is significantly different
from the value 2.34(4)  determined by
Huckestein and Kramer \cite{ref:locrefs} from the spatial decay of the
Schr\"odinger Green's function.
A similar analysis for the other impurity densities yields
$\nu_\xi=1.00(2)$,1.64(5),1.67(5) for $\rho_{\rm i}=2$,8,16 respectively.

These estimates suggest that $\rho_{\rm i}$ is associated with a weakly
irrelevant operator responsible for a slow crossover to
the ``quantum percolation'' fixed point \cite{ref:quantperc}, perhaps
flowing from the classical percolation fixed point.
In an attempt to investigate the issue of the correlation length exponent
more closely we were led to calculations of the
disorder-averaged absolute value of the antipodal
Green's function, $G_e(\nu,S)$, for the same systems. Here we can only
summarize the
findings, which will be reported in detail elsewhere.
At $\rho_{\rm i}=4$ and for system sizes up to $2S=241$, we find that
there is excellent data collapse (see Fig.~4) with $\nu_\xi = 1.64(3)$.
For the remaining impurity densities we estimate $\nu_\xi=1.38(3)$,2.02(7),
2.16(8) for $\rho_{\rm i}=2$,8,16
respectively. As the two sets of estimates for $\nu_\xi$, from $G_R$ and
$G_e$, show the same
systematics we do not view the difference between them as significant
\cite{ref:blah}.

A comment on the stability of our estimate for $\eta$ is in order. This
stability is somewhat surprising given the crossover effects in our estimates
of $\nu_\xi$. One possible explanation relies upon the numerically plausible
identification of $\eta$ with the fractal dimension
of the critical eigenstates which has been estimated to be approximately
1.6 \cite{ref:fractal} and
is not greatly different from the value $1.75$ for classical percolation. If
the latter is indeed the point of departure for the crossover the robustness
of $\eta$ becomes plausible.

To recapitulate, we have established that Read's operator does define a
legitimate order parameter for the $\sigma_{xy}=e^2/h$ disordered quantum Hall
system and by analogy, for other quantum Hall systems. At the transition
between the ordered ($\sigma_{xy}=e^2/h$) and disordered ($\sigma_{xy}=0$)
phases, we find that the ODLRO of the Read bosons exhibits a scaling behavior
consistent with that of the one-electron properties. It is worth noting that
the ordered phase is, in terms of the bosonic description, a superfluid state
that resembles a vortex glass while the disordered phase resembles a bose
glass; hence, there is a very suggestive analogy to the field tuned transition
in dirty superconductors as discussed by Fisher \cite{ref:mpaf1} in his
treatment of the latter \cite{ref:holeop}.

Finally, let us note some speculative implications for a couple of related
issues. First, the consistent
scaling of the ODLRO and Green's function implies, roughly
speaking, that multiplication of the wavefunctions of the disordered Bose
problem by the Jastrow factor $\prod_{ij} (z_i - z_j)$ is innocuous (at least
as far as the correlation length exponent is concerned). This appears to give
support to the claim of Jain, Kivelson and Trivedi \cite{ref:JKT} that their
``composite fermion'' wavefunctions for the fractional QHE yield transitions
in the same universality class as those for the integer QHE.
Second, the GM/ZHK bosonization differs from Read's in that the former
involves only the phase of the above Jastrow factor and that the resulting
ODLRO is algebraic, even in the ideal states. Our cluster interpretation
of the Read correlator results suggests that the GM
correlator in the disordered phase will be characterized by the same
$\nu_\xi$. However, it is less clear whether the GM correlator exhibits the
{\em same} algebraic decay thoughout the ordered phase, and we believe that
this remains an interesting topic for future work.

{\it Acknowledgments} --- We have benefited from conversations
with M. R. Bradley, M. E. Fisher, E.H. Fradkin, P. M. Goldbart,
D. A. Huse, S.A. Kivelson, M. P. Lombardo, M. Ma, F. C. Zhang and S. M. Girvin
who was also kind enough to comment on the manuscript. This work was supported
in part by NSF grant Nos. DMR 91--22385, DMR 91--57018 and the
Aspen Center for Physics (SLS), and the MacArthur Chair at the University of
Illinois (MPG). We also acknowledge the Materials Research Laboratory at UIUC
for use of their computational facilities.

% REFERENCES

%figures

\begin{figure}
\caption{
The antipodal Read correlator, $G_R(\nu,S)$, plotted versus $\nu$ for
$\rho_{\rm i}=4$ at $2S=20$, 60, and 180 (from top to bottom).
The data are averages over 3600 samples; the statistical errors
are less than 1\%.
Inset: $G_R(\nu,S)$ plotted versus $2S$ for $\nu=0.9$, 0.5, and 0.1,
from top to bottom.  Note that the $\nu=0.9$ data saturates at large $S$,
while at $\nu=0.1$ it vanishes exponentially with $S^{1/2}$.
}
\label{fig1}
\end{figure}

\begin{figure}
\caption{
Double logarithmic plots of $G_R(1/2,S)$ ($\times$), $dG_R/d\nu(1/2,S)$
($\Diamond$), and their ratio ($+$), versus $2S$. The line is a least-squares
fit to the data for the ratio ($2S=40$ through 320); the latter have been
shifted vertically to accommodate them on the plot.
}
\label{fig2}
\end{figure}

\begin{figure}
\caption{
Scaling plot, of $G_R(\nu,S)/G_R(1/2,S)$ versus $S^x(\nu-1/2)$
at $\rho_{\rm i}=4$, for $2S=20$, 40, 80, 100, 140, 180, 240,
320 and $|\nu-1/2|<0.1$.
The continuous curve is the data at $2S=320$.
The value of $x$ used is 0.342, corresponding to $\nu_\xi=1.46$.
}
\label{fig3}
\end{figure}

\begin{figure}
\caption{
Scaling plot of the Green's function, $G_e(\nu,S)$, data at $\rho_{\rm i}=4$
for all $\nu$ and $2S=41$, 61, 81, 101, 141, 181, 241.
The ordinate is $(\ln [G_e(1/2,S)/G_e(\nu,S)])^{1/2}$ and the abscissa is
$S^x(\nu-1/2)$ with $x=0.31$ corresponding to $\nu_\xi=1.6$.
Note that $G_e(\nu,S)=G_e(1-\nu,S)$.
The different curves correspond to different $S$ values. The data
are plotted as curves rather than as points for clarity; the error
bars are smaller than the width of the curves except for points near
the origin.
}
\label{fig4}
\end{figure}


\begin{references}

\bibitem{ref:bible} For a general introduction see R. E. Prange and
S. M. Girvin eds, {\em The Quantum Hall Effect}, Springer-Verlag, New York
(1990).

\bibitem{ref:gm} S. M. Girvin and A. H. MacDonald, Phys. Rev. Lett. {\bf 58},
1252 (1987).

\bibitem{ref:zhk} S.-C. Zhang, T. H. Hansson and S. Kivelson, Phys. Rev. Lett.
{\bf 61}, 82 (1989).

\bibitem{ref:read} N. Read, Phys. Rev. Lett. {\bf 61}, 86 (1989).

\bibitem{ref:rh-read} E. H. Rezayi and F. D. M. Haldane, Phys. Rev. Lett.
{\bf 61}, 1985 (1988).

\bibitem{ref:lkz} The corresponding scenario in the GM/ZHK scheme was discussed
by D.-H. Lee, S. A. Kivelson and S.-C. Zhang, Phys. Rev.  Lett. {\bf 67},
3302 (1991).

\bibitem{ref:locrefs} Scaling at the transition was established by
B. Huckestein and B. Kramer, Phys. Rev. Lett.  {\bf 64}, 1437 (1990). For
subsequent work see e.g. Y. Huo, R. E. Hetzel and R. N. Bhatt,
Phys. Rev. Lett. {\bf 70}, 481 (1993); and D. Z. Liu
and S. Das Sarma, Phys. Rev. B {\bf 49}, 2677 (1994) and references therein.

\bibitem{ref:sphere} F. D. M. Haldane, Phys. Rev. Lett. {\bf 51}, 605 (1983);
G. Fano, F. Ortolani and E. Colombo, Phys. Rev.  {\bf B 34}, 2670 (1986).

\bibitem{ref:fn1} The choice of the bosonic operator in both the GM/ZHK and
Read schemes is specific to the state under consideration, {\it e.g.,} for the
$\nu=2/5$ state two electrons are created following the insertion of five
flux quanta.

\bibitem{ref:laugh-gauge} R. B. Laughlin, Phys. Rev. B {\bf 23}, 5632 (1981);
B. I. Halperin, Phys. Rev. B {\bf 25}, 2185 (1982).

\bibitem{ref:thouless} Y. Gefen and D. J. Thouless, Phys. Rev. B {\bf 47}
10423, (1993).

\bibitem{ref:vortglass} W. Y. Shih, C. Ebner and D. Stroud, Phys. Rev. B
{\bf 30}, 134 (1984); M. P. A. Fisher, Phys. Rev. Lett. {\bf 62}, 1415 (1989).

\bibitem{ref:fn3}  A somewhat puzzling point is that for different
$\rho_{\rm i}$ the curves of $G_R$ versus $\nu$ at the same $S$ exhibit
crossings very close to half filling, even for $2S$ as small as 20.
Indeed, the entire distribution function
for the absolute-value of the antipodal correlations appears to be universal
close to $\nu=1/2$.
This is puzzling only because we are manifestly  {\em not}
in the scaling regime.

\bibitem{ref:quantperc} J. T. Chalker and P. D. Coddington,
J. Phys. C {\bf 21}, 2665 (1988); D.-H. Lee, Z. Wang, and
S. Kivelson, Phys. Rev. Lett. {\bf 70}, 4130 (1993).

\bibitem{ref:blah} Two remarks are in order. At a fixed impurity density
a generic renormalization group scenario would require that different
operators define correlation lengths that diverge with the same exponent
which strengthens our belief that the observed discrepancy is a crossover
effect. However, we have {\em not} observed the crossover directly as a
function of system size. We are currently extending our studies to larger
system sizes in an attempt to verify this directly.

\bibitem{ref:fractal} The critical eigenstates are multifractal; hence the
relevant quantity is the generalized dimension $D(2)$, see W. Pook and
M. Janssen, Z. Phys. B {\bf 82}, 295 (1991) and H. Aoki, Phys. Rev. B {\bf
33}, 7310 (1986).

\bibitem{ref:mpaf1} M. P. A. Fisher, Phys. Rev. Lett. {\bf 65}, 923 (1990).

\bibitem{ref:holeop} One can also define a Read operator for the holes
(vortices), $\phi^{\dagger}_{H}({\bf x};S)= F({\bf x}; S)$, which condenses
for $\nu < 1/2$ and vanishes for $\nu > 1/2$. This dual description of the
transition is a feature of any QH transition and also of the purely bosonic
description of the vortex glass/insulator transition (previous reference).

\bibitem{ref:JKT} J. K. Jain, S. A. Kivelson, and N. Trivedi,
Phys. Rev. Lett. {\bf 64}, 1297 (1990); {\it op. cit.}, 1993 (E)
(1990).

\end{references}
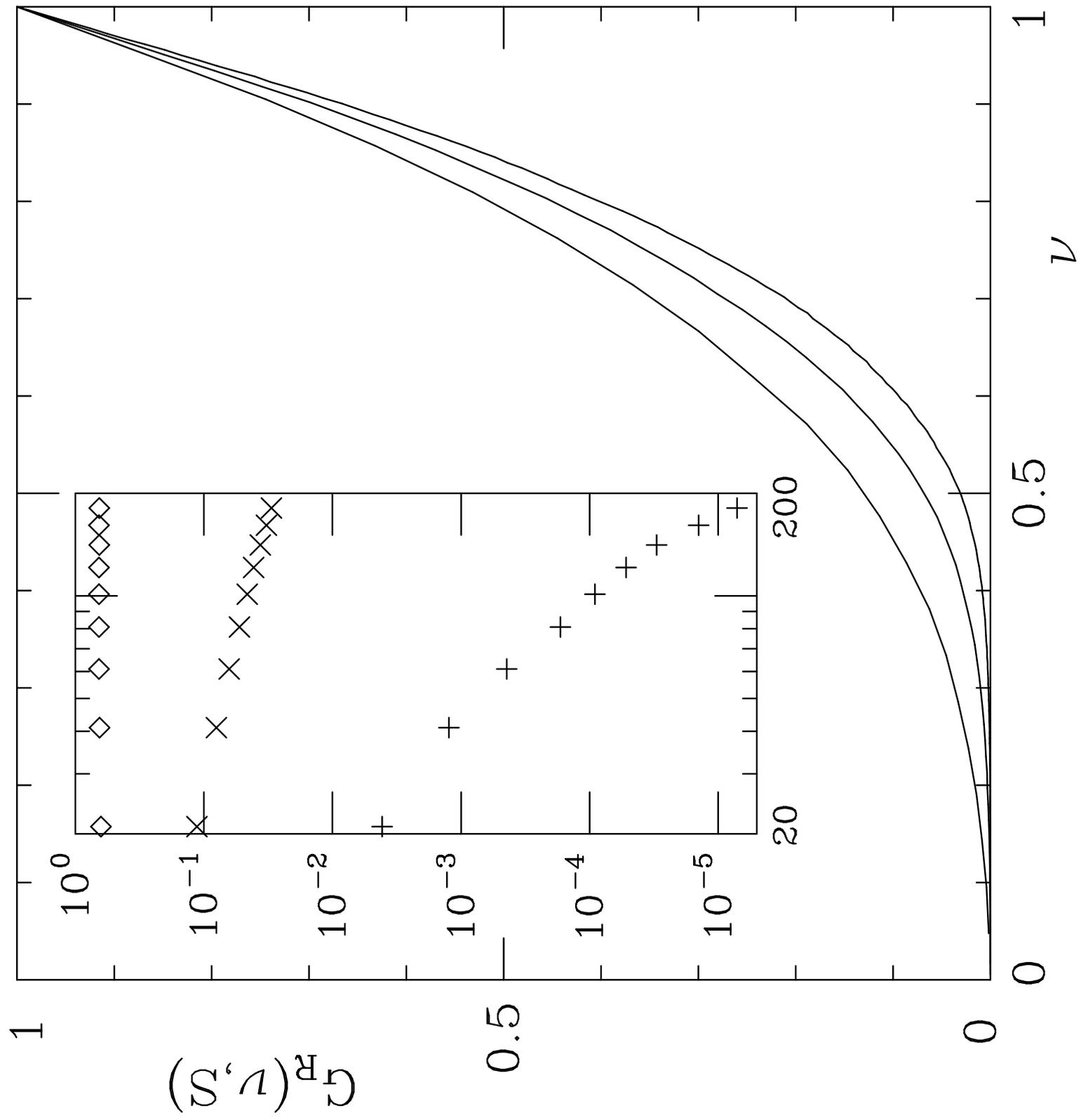
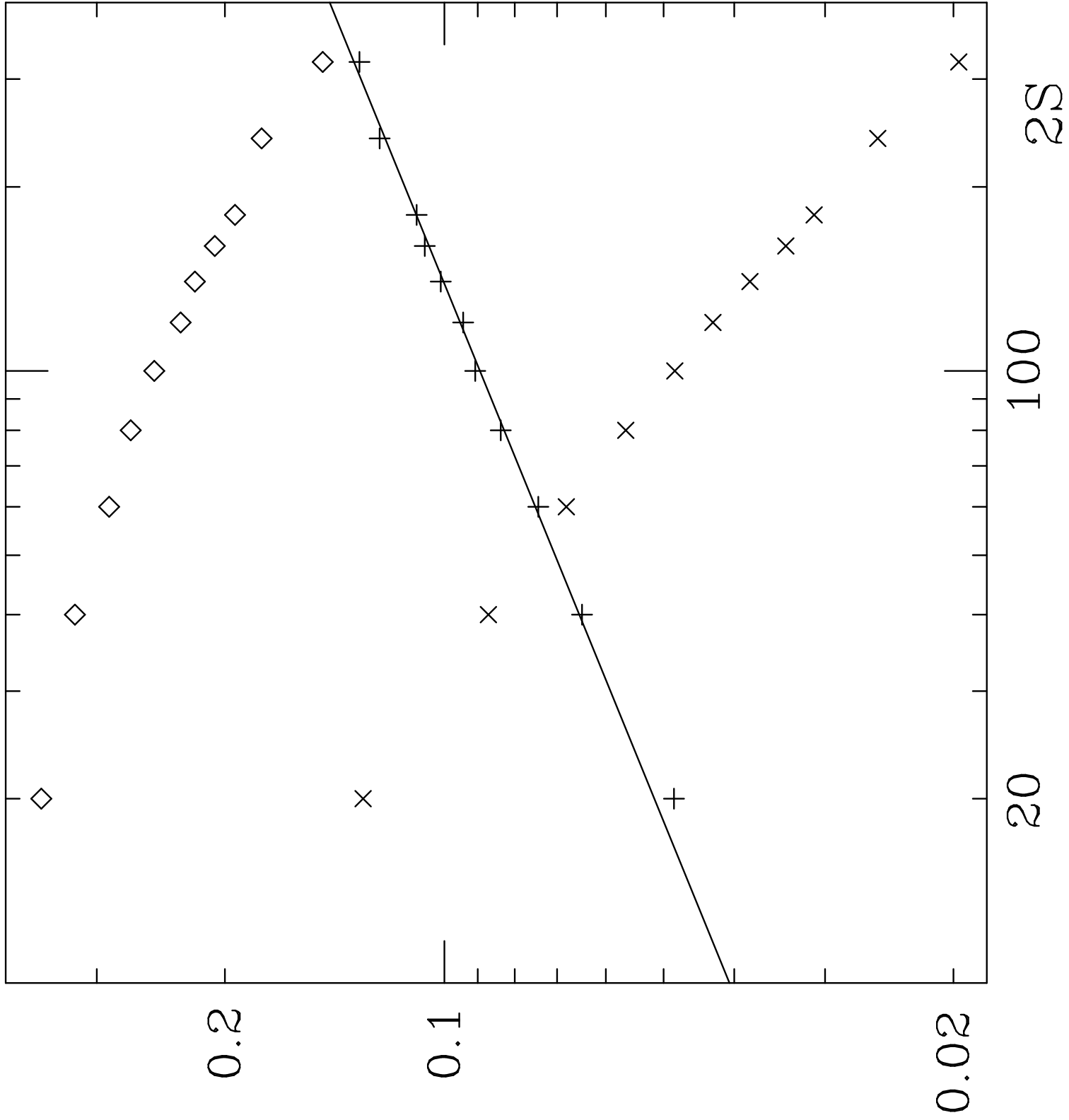
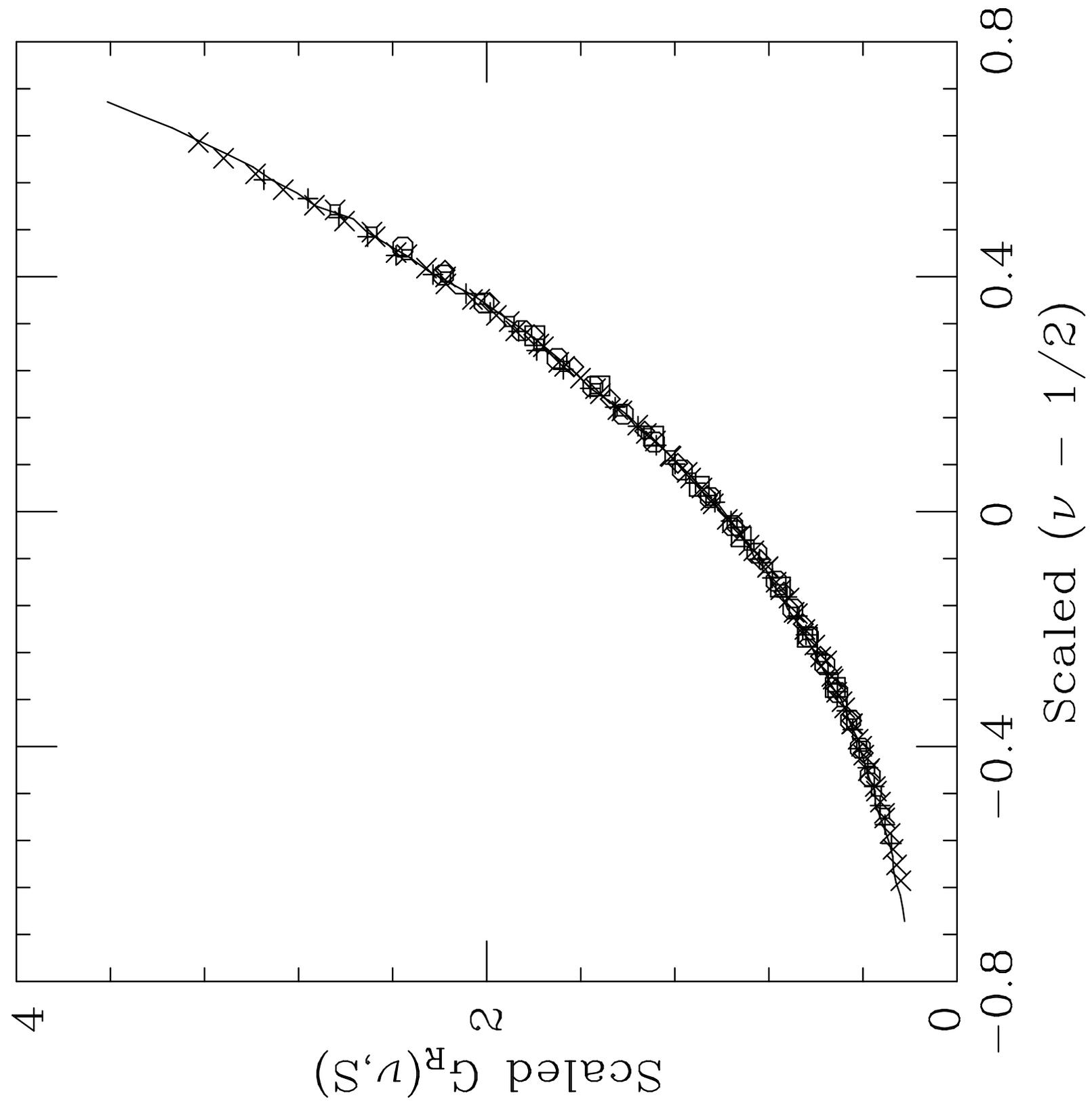
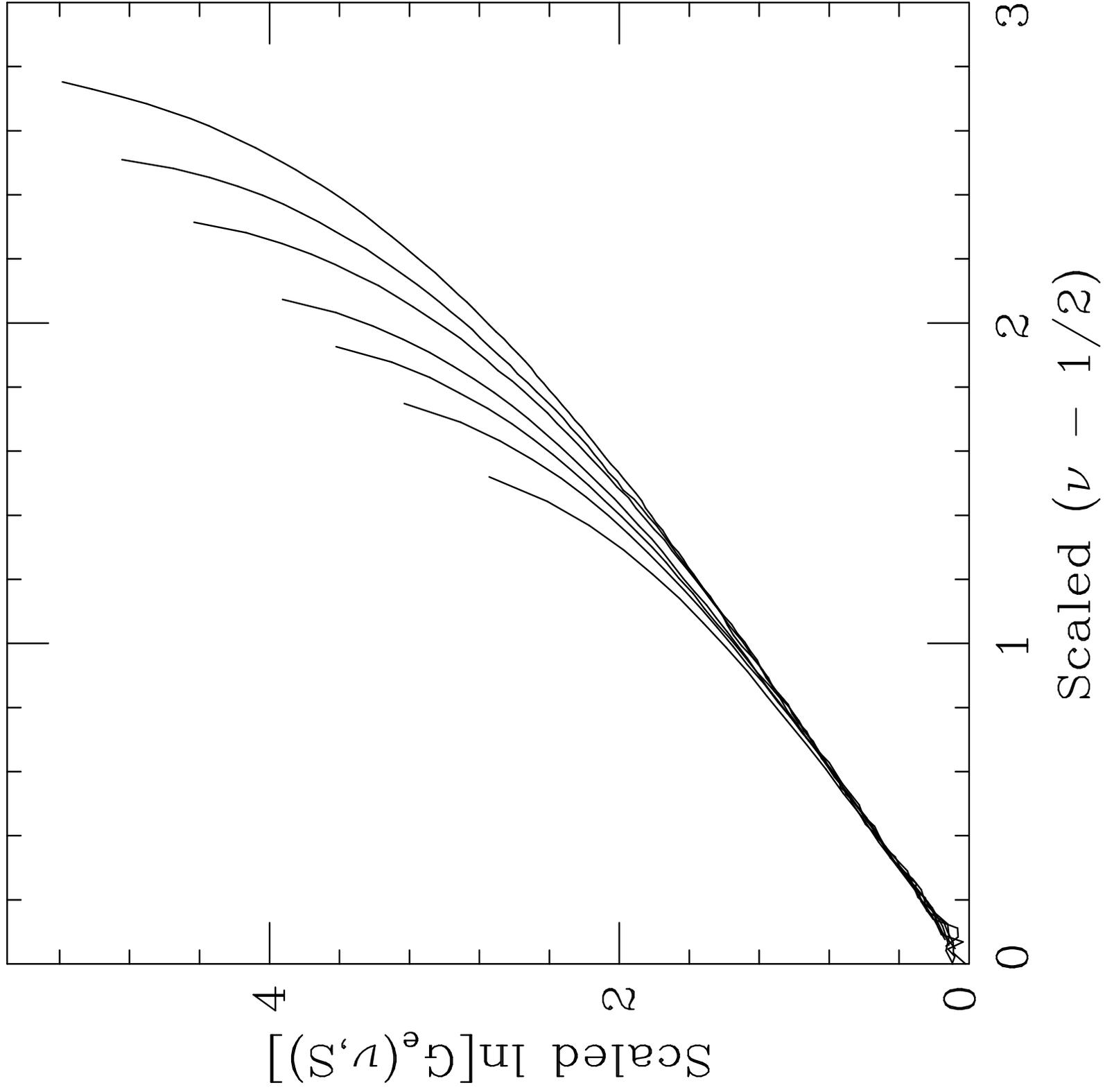
\end{document}